\documentclass[doublespacing]{elsart}
\usepackage{stmaryrd}
\usepackage{amsmath}
\usepackage{amssymb}
\usepackage{amsfonts}
\usepackage{graphicx}
\date{}
\begin{document}

\noindent \textbf{Corresponding author: }\\
 \noindent Dr. Hong-Jian Feng\\
\noindent Room 324, nineteenth Dormitory,\\
XueYuan Road No.37 ,HaiDian District,\\
Beijing University of Aeronautics and
Astronautics,\\ Beijing 100083, People's Republic of China\\
 Tel.:
+86-10-82328462\\
 Fax: +86-312-3620575
 \\
 Email address:
fenghongjian@ss.buaa.edu.cn\\
 fenghongjian@126.com
\clearpage

\begin{frontmatter}



\title{First-principles prediction of coexistence of magnetism and ferroelectricity in rhombohedral Bi$_2$FeTiO$_6$}


\author{Hong-Jian  Feng , Fa-Min Liu }

\address{Department of Physics, School of Science, Beijing
University of Aeronautics and Astronautics, Beijing 100083, P. R.
China}

\begin{abstract}
First principles calculations based on the density functional theory
within the local spin density approximation plus U(LSDA+U)scheme,
show rhombohedral Bi$_2$FeTiO$_6$ is a potential multiferroic in
which the magnetism and ferroelectricity  coexist . A ferromagnetic
configuration with magnetic moment of 4 $\mu_B$ per formula unit
have been reported with respect to the minimum total energy.
Spontaneous polarization of 27.3 $\mu$ C/cm$^2$ ,caused mainly by
the ferroelectric distortions of  Ti, was evaluated using the berry
phase approach in the modern theory of polarization. The Bi-6s
stereochemical activity of long-pair and the `d$^0$-ness' criterion
in
  off-centring of Ti  were coexisting in the predicted new system. In view of the
oxidation state of  Bi$^{3+}$,Fe$^{2+}$,Ti$^{4+}$, and O$^{2-}$ from
the orbital-resolved density of states of the Bi-6p, Fe-3d,Ti-3d ,
and O-2p states,the valence state of Bi$_2$FeTiO$_6$ in the
rhombohedral phase was found to be
Bi$_2$$^{3+}$Fe$^{2+}$Ti$^{4+}$O$_6$.

\end{abstract}

\end{frontmatter}


\section{ Introduction}
Multiferroic materials have coupled magnetic,electric,and even
structural order parameters that
 result in ferromagnetic, ferroelectric, and/or ferroelastic properties at same phase.The magnetization
can be rotated or even reversed by the ferroelectic reversal, and
vise versa. These materials have potential applications in
information storage, the emerging field of spintronics,and sensors.
Since 1960s various
 articles have systematically classified their properties and
behaviors \cite{1,2,3,4,5}, and the theoretical study was also
carried out on the magnetoelectric coupling effect\cite{6,7}.
BiFeO$_3$  was the rare one in nature, which possess both
 weak ferromagnetism and ferroelectric characteristics in single phase.
 Meanwhile, more experimental investigation have been performed on the
magnetoelectric properties of it . BiFeO$_3$ has long been known to
be ferroelectric with a Curie temperature of about 1103 K and
antiferromagnetic(AFM) with a N\'{e}el temperature of 643 K. The Fe
magnetic moments were coupled ferromagnetically in the (111) plane
and antiferromagnetically in the adjacent plane along [111]
crystalline direction, which is called the G-type AFM. The
rhombohedral distorted perovskite structure with space group $R3c$
permits a canting of AFM sublattice, resulting in a weak
ferromagnetism. However, there is a spiral spin  structure in which
the antiferromagnetic axis rotates through the crystal with a
 long-wavelength period of 620{\AA} \cite{8}.
Therefore, the linear magnetoelectric coupling was destroyed  with
the cancellation of magnetization, due to this long spiral spin
structure. Although this cancellation could be suppressed partly in
thin film , the magnetization observed in the BiFeO$_3$ thin film
was relatively smaller in terms of the excellent ferroelectric
behavior.

This cancellation of magnetization might be inhibited by doping
foreign magnetic transitional metal  ions such as Cr,Mn,Co,and Ni in
the B sites. BiCrO$_3$ was recently synthesized and reported to be a
highly distorted perovskite with $C2$ symmetry\cite{9},which permit
the occurrence of ferroelectricity. Unfortunately ,
antiferromagnetic ordering was reported with a residual
magnetization of 0.02 $\mu_B$,suggestive of weak ferromagnetism.
Bi$_2$FeCrO$_6$ has been predicted theoretically to be
ferrimagnetic(2 $\mu_B$ per formula unit)and
ferroelectric($\backsim$80 $\mu C/cm^2$).The predicted ground state
structure is very similar to the $R3c$ structure of BiFeO$_3$,except
that the Fe$^{3+}$ in the second (111) plane are replaced by the
Cr$^{3+}$, which lead to the space group $R3$.

It is known that the ferroelectricity in BiFeO$_3$ is caused by the
long Bi-6s stereochemically  active pair, which formed the one
chemical mechanism for stabilizing the ferroelectricity . This
mechanism was caused by the mixing between the $(ns)^2$ ground state
and  a low-lying $(ns)^1(np)^1$ excited state, which can only occur
if the cation ionic site does not have inversion symmetry. The
second mechanism leading to the polarization in ferroelectrics is
the off-centre displacement of small cation in the common perovskite
ferroelectrics such as BaTiO$_3$, which is attributed to the
ligand-field hybridization of the transitional metal cations by its
surrounding anions. This mechanism requires the `d$^0$-ness'
criterion in the small transitional metal off-centring, hence
precludes the coexistence of magnetism and ferroelectricity.
Sometimes these two mechanisms occur in the same material ,such as
PbTiO$_3$.

Due to the theories mentioned above, the Ti-doping BiFeO$_3$ could
maintain these two behaviors together in same phase and generate the
coexistence of magnetism and ferroelectricity. The new modified
material Bi$_2$FeTiO$_6$ with rhombohedral structure was expected to
have excellent magnetic and ferroelectric properties. As in
Bi$_2$FeCrO$_6$, the Fe$^{3+}$ in the second (111) plane are
replaced by the Ti$^{4+}$, which lead to the space group $R3$. Up to
now, there is no theoretical prediction on this new system. In this
paper we have systematically investigated the electronic properties,
ferroelectric behaviors, and magnetic ordering of the new
theoretically predicted Bi$_2$FeTiO$_6$ ,using first principles
calculation based on the density functional theory.

 The remainder of this article is structured as
follows:In section  \uppercase\expandafter{\romannumeral2}, we
presented the computational details of our calculations.  We
provided the calculated results and discussions in section
 \uppercase\expandafter{\romannumeral3}.
   In section \uppercase\expandafter{\romannumeral4} , the
conclusion based on our calculation were given.

\section{ Computational details}
\subsection{ Methodology}
 Calculations in this work have been done using the
Quantum-ESPRESSO package\cite{10}, which is based on the density
functional theory(DFT) and density functional perturbation
theory(DFPT) using the plane-wave pseudopotential formalism. We used
our self-interaction-corrected ultrasoft pseudopotential
implementation with the Perdew Burke Ernzerhof (PBE) exchange
correlation functional, as the common local density approximation
fail to obtain a band gap in the transitional metal oxides. In our
DFT computations, plane-wave basis set with kinetic energy cut-off
of 40 Ry was employed, and it has shown that the results are well
converged at this cutoff. Bi 5d ,6s, and 6p electrons,Fe 3s,3p, and
3d electrons, Ti 3s, 3p, and 3d electrons,and O 2s and O 2p
electrons have been treated as valence states. We used up to
$6\times6\times6$ grids of special k points in total energy
calculation.   In the local spin density approximation plus
U(LSDA+U) framework, the strong Coulomb repulsion is treated by
adding a Hubbard-like term to the effective potential, leading to an
improved description of the interaction in the transitional metal
oxides. The Hubbard parameter U in the range 0 eV - 8 eV were used
in the calculation of electronic structures. Results show that the U
of 7 eV is enough to produce a better insulated solution, which is
used in the following sections.

\subsection{ Structure optimization}
 The lattice parameters  was evaluated
 with respect to the lowest total energy ,ranging from
5.409 to 5.549 {\AA}. In this process the structure was constrained
to be rhombohedral within $R3$ space group. The  atomic positions
were taken from the corresponding value of BiFeO$_3$ in
Ref.\cite{11} , as the Fe$^{3+}$ in the second (111) plane are
replaced by the Ti$^{4+}$ . The calculated lattice constant value
was plotted with respect to energy and fitted to the Murnaghan
equation of states ,which give the bulk modulus\cite{12}. The atomic
positions were relaxed afterward by minimization of the
Hellman-Feynman forces within a convergency threshold of 10$^{-3}$
Ry/Bohr , using Broyden-Fletcher-Goldfarb-Shanno(BFGS) variable
metric method, as cell shape and volume was fixed.

We doubled the size of the unit cell along one of the rhombohedral
lattice vectors and compared the total energies for all possible
collinear spin configurations. The ferromagnetic(FM) and
antiferromagnetic(AFM) structure (with nearest neighbor Fe
antiparallel to each other) were constructed within the resulting
supercell up to 20 atoms. Calculated energy show that the AFM
structure supercell possess a relatively high energy of 0.03 eV per
unit cell in comparison with the FM one. The FM ordering along z and
x(y) direction were considered to find the stable FM configuration.
Total energy per formula unit show the FM ordering along z gave the
stable magnetic configuration. Hence, in the following computation
 we took the FM ordering along z as the stable state .

\subsection{ Spontaneous polarization}
In the modern theory of polarization approach\cite{15,16,17}, the
total polarization \textbf{P} for given crystalline geometry can be
calculated as the sum of ionic and electronic contributions. The
ionic contribution is calculated by summing the product of the
position of each ion in the unit cell with the nominal charge of its
rigid core. The electronic contribution is determined by evaluating
the phase of the product of overlaps between cell-periodic Bloch
functions along a densely-sampled string of neighboring points in
k-space.

\section{ Results and discussion}

The optimization of volume  was performed with lattice constant
ranging from 5.409 to 5.549 {\AA}, and the volume plotted versus
energy was shown in Fig.1.  The calculated value was fitted to the
Murnaghan equation of states, permitting to evaluate the bulk
modulus. The normal parabolic curve gave the optimal value of
5.506{\AA} ,compared with the corresponding value of 5.459{\AA} for
 BiFeO$_3$\cite{11}. We also reported the magnetic moment per formula
 unit versus volume so as to see the variation of it with respect to
 volume change, and the optimal  volume give a magnetic moment of 3.99 $\mu_B$
 per formula unit. Next, the atomic positions were relaxed using the
 optimal lattice constant with cell shape and volume fixed. The relaxing Wyckoff positions,
 calculated lattice constant
 ,rhombohedral angles, and unit cell volumes  were all summarized
in Table \uppercase\expandafter{\romannumeral1}.The internal
structural parameters are very similar to those in BiFeO$_3$ ,except
that the O positions were deviated from the initial positions. This
may partly reflects the fact that there were change of hybridization
in the new predicted Bi$_2$FeTiO$_6$.

For the simple system consisting of N atoms, transforming the total
Humiltonian into Wannier representation for the narrow band gap
condition we get
\begin{equation}
H = \sum_{i,j}\sum_\delta T_{ij} C_{i\delta}^+C_{j\delta}+
\frac{1}{2}\sum_{i,j,l,m}\sum_{\delta, \delta^\prime} <ij|v|lm> C_{i
\delta}^+C_{j \delta^\prime}^+C_{m \delta^\prime}C_{l \delta}
\end{equation}
The second term that is the interaction part can be simplified with
considering the short distance effect on one lattice or atom as:

\begin{equation}
\frac{1}{2} U \sum_i \sum_{\delta,\delta^\prime}
C_{i\delta}^+C_{i\delta^\prime}^+C_{i \delta^\prime}C_{i
\delta}=\frac{1}{2} U \sum_i\sum_\delta n_{i\delta}n_{i
\bar{\delta}}
\end{equation}

$n_{i\delta}$ and $n_{i \bar{\delta}}$ is the operator of spin-up
and spin-down number of particles.

Here U is

\begin{equation}
<ii|v|ii>=e^2 \int \frac{a^\ast (\mathbf{r-R_i}) a^\ast
(\mathbf{r^\prime - R_i}) a (\mathbf{r-R_i})
a(\mathbf{r^\prime-R_i})}{|\mathbf{r - r^\prime}|} d \mathbf{r} d
\mathbf{r^\prime}
\end{equation}

Hence, U is the Coulomb interactions between the electrons around
the lattice or atom and take great effect on the metal-insulator
transition. The electronic structures were calculated with U varying
from 0 eV to 8 eV. The results show the U of 7 eV  is able to give a
better insulated solution, while the insulated characteristics have
not been  improved significantly after increasing of U further. We
used this value of U in the following sections. The total density of
states(DOS) for Bi$_2$FeTiO$_6$ before and after applying U was
shown in Fig.2. It can be seen the band gap was increased from 0.46
eV (U=0 eV)to 1.68 eV(U=7 eV). The top valence bands caused by Ti
was pushed to the conduction bands after employed the U of 7 eV.

In order to gain more insight into the role of ferroelectric
distortions on electronic structure and bonding behavior in these
new predicted Bi$_2$FeTiO$_6$. The calculated band structure of
up-spin states was given in Fig.3. There are strong hybridization
between Bi-6p and O-2p states in the top of the valence bands, and
the Fe-3d states was pushed down in lower energy region compared
with those in BiFeO$_3$. The strong hybridization also reflect the
fact that the  activity of Bi-6s long pair also take place in this
new $R3$ structure. The band at -10 eV  are mainly originating from
the Bi-6s electrons ,these two bands are relative broader in this
new system. Ti-3d bands mainly positioned at the bottom of
conduction bands and hybridized deeply with O-2p states in this
region, and the Ti 3d also overlap slightly with O 2p  under the
Fermi energy.

To understand the change of hybridization of Ti-O and Fe-O after Ti
doping , the calculated charge density in the energy window from
$E_F$-6 eV to $E_F$ covering the region ,where the Ti-O bonding
states are located ,were shown in Fig.4 and Fig.5,respectively. From
these two Figures, it is clear that Ti is bonding strongly with O
,while Fe-O bonding strength is reduced comparing with those in
BiFeO$_3$. Moreover the Ti-O bonding is along the direction in which
the Ti-3d orbitals formed and slightly affected by the ligand-field.
It means the Ti is bonding deeply with the nearest six Oxygen atoms.
Therefore the Ti off-centring driving ferroelectric displacement
still maintained in these new system. We are expecting this chemical
mechanism for stabilizing the distorted structure in ferroelectric
oxides take effect in the new system.

The Ti tend to form the oxidation state of Ti$^{4+}$, which can be
approved by the orbital-resolved DOS for Ti shown in Fig.6
consequently. It is found the $d_{xy},d_{yz},d_{z^2},d_{xz}$,and
$d_{x^2-y^2}$ orbitals ,including the majority spins and minority
spins, are all nearly unoccupied in the conduction bands ,indicating
the oxidation state of Ti$^{4+}$. These bands are hybridizing
strongly with the O-2p electrons, leading to the stabilizing
ferroelectric off-centring behavior. And this phenonmenon can
further confirm the occurrence of this chemical mechanism in this
predicted new system. The $d_{yz},d_{xz}$,and $d_{x^2-y^2}$ orbitals
have rather little states broadened under the Fermi energy, which
indicate the smaller deviation of oxidation state from Ti$^{4+}$.
These bands also overlap with O-2p electrons to strengthen the
hybridization.

In the new structure, Fe atom tend to form oxidation state of
Fe$^{2+}$ ,which is lowered comparing with the Fe$^{3+}$ in
BiFeO$_3$. We reported the orbital-resolved DOS for Fe in BiFeO$_3$
and Bi$_2$FeTiO$_6$   shown in Fig.7 and Fig.8 respectively , so as
to observe the change of oxidation states. It can be seen in Fig. 7
that all the majority spin states are occupied while all the
minority spin states are nearly unoccupied with fewer states lying
at -3.5 eV ,indicating the  oxidation state of Fe$^{3+}$ in
BiFeO$_3$. However ,from orbital-resolved DOS of Fe in
Bi$_2$FeTiO$_6$ shown in Fig.8 , unlike the case in BiFeO$_3$ , the
spin-down states in $d_{xy}$ and $d_{xz}$ orbitals are partly
occupied ,indicating the orientation of bonding in this
circumstance. Meanwhile all the spin-up states are occupied as in
BiFeO$_3$. Based on the large amount of occupied spin-down states in
$d_{xy}$ orbitals and the relatively fewer occupied spin-down states
in $d_{xz}$ orbitals, we deduced the formal spin (S=4/2) of
Fe$^{2+}$ with $d^6$ configuration in Bi$_2$FeTiO$_6$. Although the
Fe-3d states were narrowed and pushed close to the Fermi energy
after doping with Ti, the main states have not overlapped deeply
with the O-2p electrons. Hence the Fe 3d-O 2p hybridization have not
been strengthened consequently.

The Bi-6s stereochemically active long pair were participating in
the ferroelectric distortions. The hybridization between Bi-6p
electrons and O-2p electrons can be observed in the orbital-resolved
DOS of Bi and O in Bi$_2$FeTiO$_6$ shown in Fig.9. We also reported
the total DOS in Fig. 9 for comparision. It is clear that the Bi-6p
states and O-2p states positioned at the same energy region and
hybridized strongly as in BiFeO$_3$. Moreover,the Bi 6p does not
change much comparing with the states in  BiFeO$_3$ , indicating the
similar oxidation states in these two systems. In view of the
oxidation states of Ti$^{4+}$,Fe$^{2+}$,and Bi$^{3+}$, the valence
state in the $R3$ Bi$_2$FeTiO$_6$ is
Bi$_2$$^{3+}$Fe$^{2+}$Ti$^{4+}$O$_6$.

The spontaneous polarization was evaluated using the berry phase
method based on modern theory of polarization.
 And Born effective charges(BECs) were estimated by computing the Cartesian components of
 the polarization with respect to the atomic displacements, ie.:

\begin{equation}
\Delta P_{\alpha}\cong\sum_{j\beta}\frac{\partial
 P_\alpha}{\partial\mu_{j\beta}}(\mu_{j\beta}-\mu_{0j\beta})=\frac{e}{\Omega}\sum_{j\beta}Z_{j\alpha\beta}^\ast\Delta\mu_{j\beta}
 \end{equation}

where $\Delta \mu_{j\beta}$ is the displacement of ion $j$ in
Cartesian direction $ \beta$, $ Z_{j\alpha\beta}^\ast$ is its BECs
tensor, and $\Omega$ is the unit cell volume.  The $\mathbf{P}$ is
calculated from berry phase method by considering a specific
structural pathway parameterized by the change in atomic
displacement connecting a centrosymmetric reference structure and
$R3$.  The displacement along [111] direction was chosen smaller
(2\% lattice constant)enough to ensure the validity of the linear
treatment in Eq.(4). The results were listed in table
  \uppercase\expandafter{\romannumeral2}.

The new predicted Bi$_2$FeTiO$_6$ was found to obtain a medium value
of spontaneous polarization  of 27.3 $\mu$ C/cm$^2$ ,comparing with
the relative large value of 59.4 $\mu$ C/cm$^2$ in BiFeO$_3$. This
result was not so good as we expected  in terms of the coexistence
of these two chemical mechanisms of ferroelectric distortions  in
the same phase. This phenomenon might be partly explained by the
anomalous large BEC of 5.82 for Ti ,leading to the dominant role of
Ti off-centring distortions in the new predicted system.The
producing polarization was similar to the value in BaTiO$_3$
\cite{21} in which the same driving mechanism applied. Moreover,the
BEC of Bi and Fe were all decreased in the new system, indicating
the decreasing ability of producing the ferroelectricity of Bi-6s
long-pair. There might be interaction between these two chemical
mechanisms for producing ferroelectricity ,which need to be explored
further. The predominant role of Ti off-centring ferroelectric
displacement generated a smaller  ferroelectricity than the Bi-6s
long-pair in terms of a collapse of 32 $\mu$ C/cm$^2$ in
Bi$_2$FeTiO$_6$  compared with that in BiFeO$_3$ . Fortunately, the
adoption of Ti in the new system inhibited the cancelling of
macrosopic weak magnetization , accompanying the oxidation state of
 Fe$^{2+}$ with $d^6$ configuration(with magnetic moment of 4 $\mu_B$ per
formula unit ). Consequently, the coexistence of ferromagnetism and
ferroelectricity can be fulfilled in the new predicted
Bi$_2$FeTiO$_6$.

\section{ Conclusion}
The cancellation of magnetization caused by the spiral spin
structure ordering in BiFeO$_3$ can be improved by doping foreign
transitional metal ions. Bi$_2$FeTiO$_6$  was predicted to be a
potential multiferroic in which a large ferromagnetism(with magnetic
moment of 4 $\mu_B$ per unit cell) and ferroelectricity coexisted.
Berry phase calculation show a medium polarization of 27.3 $\mu$
C/cm$^2$ in Bi$_2$FeTiO$_6$ comparing with the value of 59.4 $\mu$
C/cm$^2$ in BiFeO$_3$, which was mainly caused by the Ti off-centre
ferroelectric distortions, although the hybridization between Bi-6p
states and O-2p states had taken place  in this new predicted system
, which was reduced after doping Ti ,accompanying  a small deviation
of BECs of Bi from the nominal one. From the orbital-resolved DOS of
the elements in the system ,the oxidation state of Bi$_2$FeTiO$_6$
was found to be Bi$_2^{3+}$Fe$^{2+}$Ti$^{4+}$O$_6$ with decreasing
of oxidation state of Fe from Fe$^{3+}$ to Fe$^{2+}$.

This work was supported by the Aeronautical Science foundation of
China (Grant Nos. 2003ZG51069).


\bibliographystyle{elsart-num}

\bibliography{Bi2FeTiO6}


\clearpage

\begin{table}[!h]

\caption{ Calculated lattice constant a, rhombohedral angle
$\alpha$, volume V, and Wyckoff parameters for Bi$_2$FeTiO$_6$. The
Wyckoff positions 1a(x,x,x) for the cations and 3b(x,y,z) for the
anions.}

\begin{center}

\tabcolsep=8pt
\begin{tabular}{@{}ccccccc}
\hline\hline
  &&Bi2FeTiO$_6$   &BiFeO$_3$&\\
\hline
a({\AA})&&  5.506& 5.459\\
$\alpha$(\textordmasculine )&&60.36&60.36\\
V({\AA$^3$})&& 118.97  & 115.98 \\
Bi& x& -0.001/0.501 & 0.000   \\
Fe/Ti& x& 0.229(Fe)/0.728(Ti) & 0.231  \\
O&x& 0.558 & 0.542  \\
&y& 0.935 & 0.943  \\
&z&  0.393  & 0.398 \\
\hline\hline
       \end{tabular}

       \end{center}
       \end{table}

\begin{table}[!h]

\caption{The polarization(P) and Born effective
charges(BECs)(Z$^\ast$) for displacements along [111] for
Bi$_2$FeTiO$_6$ and BiFeO$_3$ .}

\begin{center}

\tabcolsep=8pt
\begin{tabular}{@{}cccccccc}
\hline\hline
& \multicolumn{3}{c}{BiFeO$_3$}&\multicolumn{4}{c}{Bi$_2$FeTiO$_6$}\\

\cline{2-4} \cline{5-8}
 P($\mu$ C/cm$^2$)   &\multicolumn{3}{c}{59.4}  &\multicolumn{4}{c}{27.3} \\
\hline
&  Bi& Fe&  O & Bi&Fe &Ti  & O\\
\hline
Z$^\ast$ &  4.41 & 3.32&  -2.58 & 3.61& 1.89& 5.82& -2.49\\

\hline\hline
\end{tabular}
\end{center}
\end{table}

\clearpage

\raggedright \textbf{Figure captions:}

Fig.1 Volume vs. total energy per formula unit and magnetic moment
per formula unit for Bi$_2$FeTiO$_6$ .

 Fig.2 (a) and (b) show total
DOS for Bi$_2$FeTiO$_6$ with U=0 eV and U=7 eV respectively.The
Fermi level was set to zero.

 Fig.3
Electronic band dispersion for Bi$_2$FeTiO$_6$. The Fermi level was
set to zero.

 Fig. 4 Charge density (in arbitrary units) in
a particular plane for BiFeO$_3$. The charge density is calculated
in the energy window from $E_F$-6 eV to $E_F$.

 Fig. 5 Charge
density (in arbitrary units) in a particular plane for
Bi$_2$FeTiO$_6$. The charge density is calculated in the energy
window from $E_F$-6 eV to $E_F$.

 Fig. 6 Orbital-resolved DOS for
Ti in Bi$_2$FeTiO$_6$.(a),(b),(c),(d),and (e) show the DOS for
$d_{xy},d_{yz},d_{z^2},d_{xz}$,and $d_{x^2-y^2}$ orbitals
respectively.Spin-up states are shown in the upper portions and
spin-down states in the lower portions in all panels.The Fermi level
was set to zero.

 Fig. 7 Orbital-resolved DOS for Fe in
BiFeO$_3$.(a),(b),(c),(d),and (e) show the DOS for
$d_{xy},d_{yz},d_{z^2},d_{xz}$, and $d_{x^2-y^2}$ orbitals
respectively.(f) show the Fe 3d total DOS.Spin-up states are shown
in the upper portions and spin-down states in the lower portions in
all panels.The Fermi level was set to zero.

 Fig. 8
Orbital-resolved DOS for Fe in Bi$_2$FeTiO$_6$.(a),(b),(c),(d),and
(e) show the $d_{xy},d_{yz},d_{z^2},d_{xz}$,and$ d_{x^2-y^2}$
orbitals respectively.Spin-up states are shown in the upper portions
and spin-down states in the lower portions in all panels.The Fermi
level was set to zero.

 Fig. 9 Total DOS and orbital-resolved DOS for Bi 6p  and O 2p in
Bi$_2$FeTiO$_6$.(a) and (b)show the Bi-6p and O-2s states
respectively. (c) is the total DOS. Spin-up states are shown in the
upper portions and spin-down states in the lower portions in (c).The
Fermi level was set to zero.

\clearpage









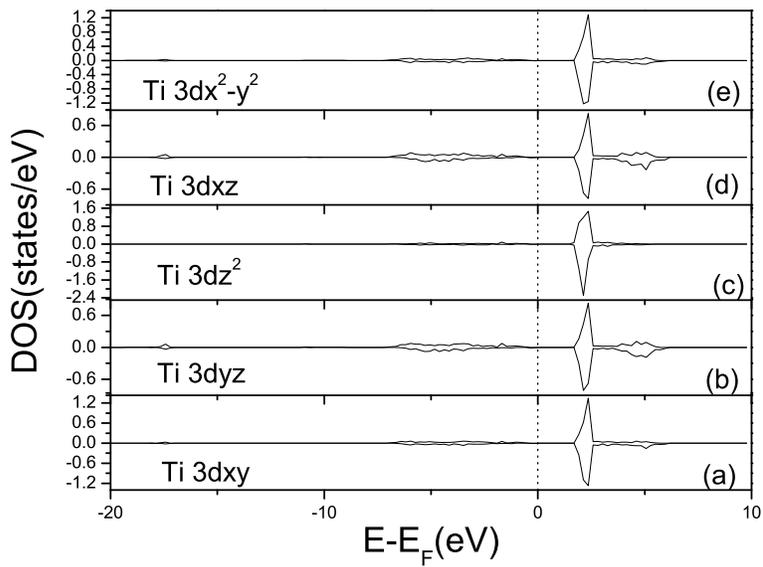
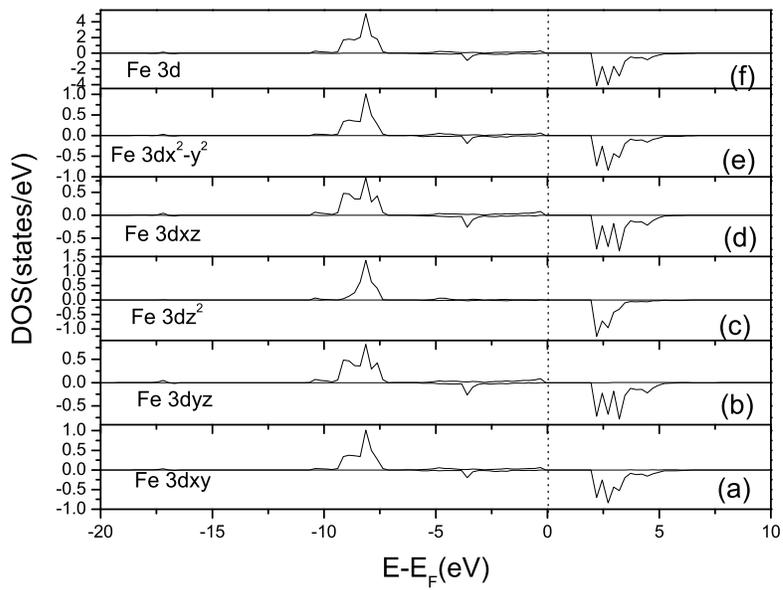
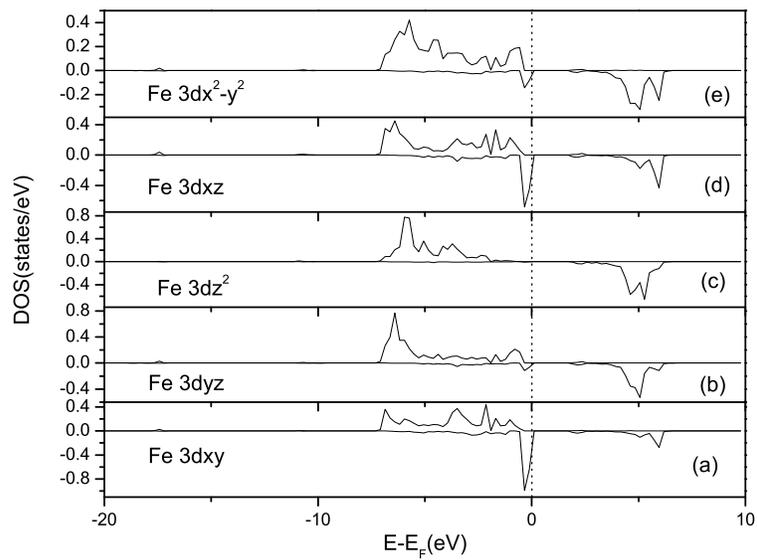
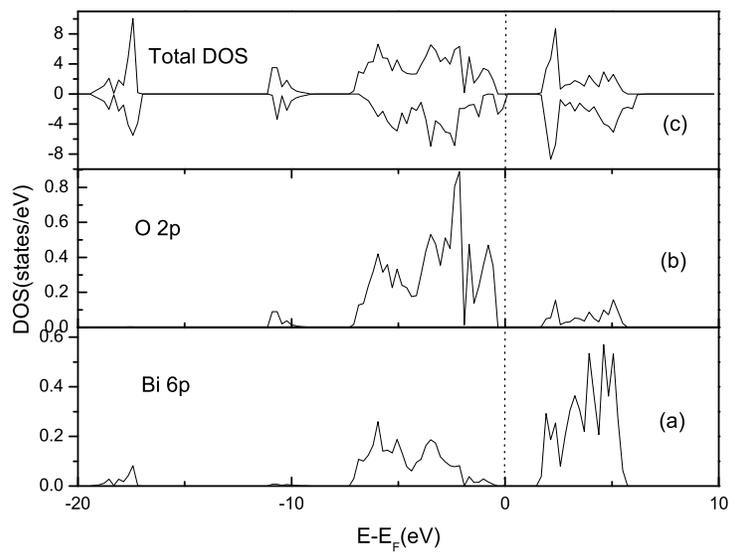
\end{document}